# A multi-model approach to building thermal simulation for design and research purposes

H. BOYER, F. GARDE, J.C. GATINA
*Université de la Réunion, Laboratoire de Génie Industriel,*
*Equipe Génie Civil, BP 7151, 15 avenue René Cassin, 97705 Saint-Denis Cedex, FRANCE (DOM)*

J. BRAU
*INSA de Lyon, CETHIL – Thermique du Bâtiment, Bât. 307, 20 avenue Albert Einstein,*
*69621 Villeurbanne Cedex, FRANCE*

Abstract

The designers pre-occupation to reduce energy consumption and to achieve better thermal ambience levels, has favoured the setting up of numerous building thermal dynamic simulation programs. The progress in the modelling of phenomenas and its transfer into the professional field has resulted in various numerical approaches ranging from softwares dedicated to architects for design use to tools for laboratory use by the expert thermal researcher. This analysis shows that each approach tends to fulfil the specific needs of a certain kind of manipulator only, in the building conception process. Our objective is notably different as it is a tool which can be used from the very initial stage of a construction project, to the energy audit for the existing building. In each of these cases, the objective results, the precision advocated and the time delay of the results are different parameters which call for a multiple model approach of the building system

INTRODUCTION

Since the initial significant diffusion of simulation and design models, one of the major problems is their appropriateness to a particular user during the design process [1,2]. According to Kendrick [3], three principle model users can be distinguished, each of which have their own particular needs:

- Building designers and operators
- Government policy makers
- Building physics researchers

For the building designers and operators, from the first stages of conception, different options should be rapidly explored to identify the important parameters. Step by step, during the progress, more detailed tools are necessary to perfect the design. Therefore, during the first phase, an architect would consider, for example, the influence of the orientation (according to the sun's course and dominant winds), or take into consideration certain shades for sun exposed walls. At a later stage, the engineering research unit is concerned with the dimensioning of the air conditioning installation and finally the building user is principally concerned by its energy consumption.

The policy maker insists that these models are capable of providing much more detailed information than that required in the building process, in that the user is concerned by the impacts on a national scale, concerning buildings.

Finally, the researcher in this field is the one responsible for providing efficient models to the first two groups. This work, whether fundamental or applied, numerical or experimental, should enable better comprehension of the physical mechanisms or better communication of recent knowledge in a professional circle. This part of his work needs to be sensitive to their needs and also aware of the limits to their knowledge in this domain.

Considering the manner in which they are constructed, the thermal models are very often obtained by a combination of elementary models, each one describing the thermal behaviour of the elements of the building. Once the elementary models have been chosen, this approach gives a rigid or *monolithic* building model. It replies precisely to the users needs (HVAC designer for example). However, in view of the very different needs of the professionals in the field, a tool of which the elementary models are fixed, will never on its own, meet the needs of all those involved in the design process.

Our contribution to this problem, by a different approach, consists of the integration of a database of elementary and interchangeable models.

On the other hand when justifying the multi-model approach, it must be made evident that the implied models are greatly conditioned by the availability of data concerning the constituents of the building. In the initial stages of a project this data is very limited (outlines, dimensions) and at each stage of the building evolution this data becomes more complex (material choice, opening locations, air conditioning systems) and thus enables the use of much greater defined models.

*Fig. 1 : Building database evolution*

For a research tool used at the final stage of the process, it is possible to obtain different levels of detail or to proceed to sensibility analysis. From the initial project outline to its realisation, the model users have very differing objectives concerning the design tool, conditioned by the preference for such or such a model. Two different objectives are for example, on the one hand to evaluate the annual energy consumption and on the other, to study the results (hourly/sub hourly) of a building component over a given climatic sequence. It must be made note that the use of a detailed thermal model in the initial case leads to considerable calculation times. Equally the use of an over simplified model will only ever give limited estimations, but ensure rapid results. A strong link exists between model quality, simulation objectives and calculation time.

**I) CERTAIN REFERENCES CONCERNING DETAILED SIMULATION CODES**

Initially the development of the simulation codes was reserved to large teams working with powerful computers, because of the complexity of the transfer modes coupling, the system sizes and the calculation times. Thus the reference codes such as ESP[4], DOE, BLAST[5] and SERI-RES were initially developed on mainframes. With the increase in power of personal computers, smaller groups were able to produce high performance codes. These powerful codes, adapted to small computers (ex [6]), are more easily diffused into the professional field. Initially these software were mostly monozone : thus for a given building (or a total of one or more rooms) the dry air temperature is taken to be the same. This is a reasonable hypothesis, when taking the rooms to have same thermal response (same exposure conditions, air conditioning and wall composition...). Examples of these tools are QUICKTEMP[7] or CODYBA[8]. The following evolution was the recognition of multizonage, which led to the first versions of software such as ESP, BLAST and COMFIE[9]. Next the realistic integration of airflow transfers posed a difficult problem. The first reason being the lack of simple models to take into account convection through large openings (doors and windows). In this case the combination of the driving effects can lead to the opposite air flows within the same opening. Since, these models have been outlined and integrated into various airflow calculation codes, as COMIS, AIRNET[10] CONTAM93[11], MOVECOMP[12] or thermal airflow codes such as TARP, ESP and our own code CODYRUN. The second difficulty is linked to the non linear character of the pressure system, of which the resolution is necessary in order to determine the mass flows. As well as the problems (convergence difficulties in particular) intrinsic to this non linear system, the setting up of the thermal airflow coupling is also problematic [13,14].

Taking into consideration the desired objectives, and the progress realised in modelling, it is to be noted that, the tools available cover a wide range, from the very simple tool used by the designer (monozone, without air flows), to the extremely detailed tool, exclusive to the building physics researcher specialist, referred to in the following paragraph.

**II) A MODEL DEDICATED TO THE BUILDING PHYSICS RESEARCHER**

In scope of the specific codes considered up to now in TRNSYS[15,16], much work has principally been carried out by research laboratories. A building is considered by taking into consideration its composition of a certain number of rooms, walls and windows. The fig. below explains the lay-out of the building, obtained from the assembly of the elementary components, such as the walls, windows, air conditioning equipements.

*Fig. 2 : Component assembly of a building*

The analytic procedure breaks down the object of the building research study into elementary parts, which is necessary to study independently, in order to understand their internal functions and interactions, in order to reconstruct the basic model of the object. Each of the elementary components, walls, glazing, air volume, can be represented (as regarding heat transfer) by various models. Thus for example, when considering a wall, the temperature field represents the balance between conduction, convection and radiation. For each of these phenomena analytic models are available and their assembly will produce the thermal model of the wall. By assembling each transfer model (conduction, convection,...) of each component ( walls, windows,...), the analytic method leads to the construction of a thermal model for the entire building. Considering the complexity of the model obtained, for even the most simplest of buildings and despite the wide range of further improvments (and the possibility of coupling various systems to the building model), it is clear that this process concerns principally the researchers within the field and is not of a priority to users such as architects.

**III ) OUR MULTIPLE MODEL SOFTWARE CODYRUN :**

As a result of communal research between the University of Reunion and INSA Lyon (*CETHIL – Building Physics Team*), this work aims to produce a high performance building thermal simulation software regrouping design and research aspects and adapted to different types of climate. More specifically, it concerns a multi-zone software including natural ventilation and humidity transfers, which is called *CODYRUN*.

The three major parts are modules describing the building, simulation and the use of the results.

In order to describe any building, we have proposed [17] to split it up into three types of entities called *zones* (according to thermal zones), *Interambiances* (separations between zones, the outside being considered as a particular zone) and *Components* (walls, windows, air conditioning systems, ...). For each of the above entities, data structures have been created regrouping information related to their description and, as explained further on, others related to associated models. For example, in a particular zone, the information about the description will be the name of the zone and its volume. The descriptive fields associated to a component will be the type of components (wall, window, air conditioning system...,), the zone number or interambiance, the surface area and the composition of the wall layers, the setting and the power for air conditioning systems... To inform all these data structures, the software was programmed under Microsoft Windows environment.

During one simulation, the possibility to offer the expert the choice between different transfer models or the reconstitution of meteorological parameters, is one of the most interesting aspects. Due to the multi-model aspect of the software it was necessary to set up various models of differing accuracy, for each of the physical phenomenon (or systems):

- Airflow transfers
- Outdoor convection
- Indoor convection
- Outdoor long wave exchanges
- Indoor long wave exchanges
- Indoor short wave repartition
- Heat conduction
- Conductive exchanges with the ground
- HVAC system

In the majority of the existing simulation software, the choice of the model is made (for each phenomena or system) during the analysis stage of the computer project, which is dependent on the objective of the software, the desired accuracy level and the calculation time. Also their application is global to the whole building (for example, an identical conduction model is applied, for all of the opaque walls). In consequence, if these models are more or less accurate, the multizone character and the consideration of the airflow transfers, leads rapidly to a calculation time which is incompatible with a conception tool. To get round this compromise between precision and calculation time, it became evident that for certain of the phenomenon (or systems), selective application must be made available for the models. This means the availability of a choice of different entity models in the same hierarchy (for example, the zones or the walls). The aim is therefore to subjugate the complexity level of the models concerning one entity, to the interest attached to that entity. Thus, for example, the choice of the conduction model carried out for wall definition level, may differ from one wall to another. The choice towards greater precision or not in the linkage, is defined by the particular interest for the wall: for example, the analysis of the condensation risks of one particular wall necessitates close precision during the calculation of the surface temperatures, whereas simpler models are more than likely adequate for the other walls of the building. From this example it appears that a selective starting process must however assure a condition concerning the intensity of the coupling of the systems at the same level (and particularly the zones) which remains the competence of the expert user.

The above point now established, it is thus necessary to give out the details informing the models in the description structures of the above entities (building, zones, components). The following table shows the allocation of the model choice which was carried out. The first line concerns the number of zones considered, which is in fact a hypothesis of the building thermal model. All the physical models found at building level are of a global character. This concerns models of airflow transfer, outdoor convection and radiation long wave exchanges. The indoor convection model, carried out at zone level, could also be carried out at wall level. However as the problems concerning the building engineer are most often linked to the thermal behaviour of one given zone, it was decided to use the same model for the whole of the interior walls and windows to one zone. An identical procedure was carried out for the indoor long wave radiation exchanges. The lowest model (as regarding our database) to which it is possible to operate, concerns the conduction in a wall (or in windows) and the split system model.

*Table 1 : model/entity allocation*

**1 ) The integrated models**

All the models concerned are not reviewed, but the interest and implications of the multiple model approach are put forward.

- *Airflow transfers*

The modelling of airflow transfers is delicate and extremely time consuming due to the non-linear character of the system which is to be solved, in order to determine the inter-rooms and outside transfers. With the addition of the wind and the thermal buoyancy, the source of the implementation difficulties lies in the coupling with the thermal model/the coupling within the thermal model. Most thermal models take these flows to be known, which only corresponds to reality in the case of a mechanically air ventilated building. Apart from these cases, regarding the open plan character of a building, especially found in a hot climate, a detailed airflow model was integrated, based on the determination at each time lapse, of the reference pressures of each zone, at each time lap. This method is used in the reference codes, as ESP or TARP. A particular point to be highlighted is the recognition of the transfers through large openings integrated in CODYRUN, through the use of the Walton model[10].

- *Sky temperature*

This fictitious temperature is introduced to model the long wave radiation exchanges with the sky. Often taken as equal to the dry air temperature of the external air, this model is faulty, especially in a temperate climate during very clear nights. Different authors offer different relations and take into consideration the external dry air temperature [18], the dew temperature, [19] or the degree of cloud cover [20]. The interest in the precise definition of this parameter became clear during the experimental validation phase of CODYRUN, during the experiments on light weight buildings with low emmisivity rooves.

- *Outdoor convection*

Taking in scope the constant exchange coefficients, many authors give relations linking the coefficient exchange value to the wind speed and direction [21-24]. The window choice of an outside convection model will be included further on in this chapter, in the section treating the link between the description and the model.

- *Indoor convection*

The interior exchange factors are the basis of many publications ([25-27]), as they finally assert the physical link between the interior walls of the envelope of the building and the vector of the principal exits (for the building heat engineer ), the dry interior air temperatures (or the air conditioning sensitive power). By means of experimental correlations, non linear relationships link the value of the exchange factors and the temperature gap between the air and the surface of the walls. If the method of integration of these models is appealing from a numerical precision point of view, these non linear relationships induce (as opposed to a linear case) a longer calculation time (as it is necessary to establish the temperature gap to be able to calculate the air and surface temperatures, and to repeat until convergence)[28].

- *Heat conduction*

The chosen method is based on the thermal and electrical analogy, as it relates well to the multi-model character of our specifications. In a preceding article [29], the principle of the building thermal model generation is explained. This generation takes into consideration the construction of a building (the choice of the number of thermal zones, wall descriptions,...) and particularly the conduction model of each wall. In theory, the identification of the transfer function of the wall may be carried out with the aid of different filters. The network of the walls, that proceeds thereof, is more or less dense, according to the desired objective. In the case of energy estimation over a long period, the R2C type model (one node to which is associated a capacity on each side of the wall) is sufficient. If, however it is necessary to analyse the risks of local condensation on the walls, a model giving more precise surface temperature estimation is necessary, for example the 3R2C model or a model networking each layer of the wall independently. Nevertheless one must keep in mind that the global thermal model of the building will have a more important dimension because the walls will be finely netted, which will be an important factor in the calculation time.

- *Indoor short wave repartition*

The most simplest model considers that the direct beam from a window is incident on the ground (or floor) and that the diffused radiation is proportionally dispersed on the surfaces. All the radiations are reflected and it is therefore necessary to establish a linear system to determine the flux densities absorbed by each interior wall according to its radiative characteristics. Considering the different geometry of any one of the zones, this system can reach great dimensions. In order to find a compromise between the two preceding processes, an intermediary strategy was adopted, inspired by CODYBA[8]. Interior data groups are formed (floors, walls, windows, interior walls/separations), each affected by mean values (balanced by the surfaces) of their radiative properties (absorption, reflection), which enables to reduce by four the dimension of the preceding system, and finally allowing the repartition of the flows on the interior real walls.

- *HVAC modelling*

We have chosen to offer different levels of model for split system type air conditioning systems. It corresponds to different objectives, such as power requirements (energetic approach) or the analysis of response of the split with a reduced time lapse. We shall look at this section in greater detail further on.

**2 )The link between the description and the model**

The link is established by taking into consideration display windows, which, classically hold the information linked to the description. Therefore, for building data structure, the association with the models is illustrated below on fig. 9. In this display window, the information is to be completed relating to the building (name, number of zones, morphology) or, as well, the implantation site (latitude, albedo,...). The buttons *Zones, InterZones* and *Components* give access to the description windows of these entities. The access to the models attached to the building entity is done through the buttons of the model section. Each of these buttons gives access to a multiple selection box. It is thus that it is possible to select the chosen models for the outdoor convection, the reconstitution of some meteorological parameter models (diffuse radiation and sky temperature) as well as the airflow model. Found to the right of each button, a text informs of the model actually chosen.

*Fig. 3 : Building description window*

Considering the multi-user character of the software, the part of screen concerning models is only accessible to experts and is invisible for the professional (it is through the different passwords available to different classes of users). An example of a model selection box is shown covering the convection exchange coefficient model for the external walls of a building.

*Fig. 4 : External convection model window*

V) CODYRUN USED AS A DESIGN TOOL

A Demand Side Management pilot initiative was launched in early 1995 in the French islands of Guadeloupe and Reunion through a partnership between the French electricity board (EDF), institutions involved in energy saving and environmental conservation (ADEME) and construction quality improvement, the ministries of Housing, Industry and the French Overseas Department, the University of Reunion island and several other public and private partners, such as low cost housing institutions, architects, energy consultants, etc... The ECODOM standard aims to simplify the creation of naturally ventilated comfortable dwellings whilst avoiding the usual necessity of a powered air cooling system consuming electricity. The aim of ECODOM is to provide simple technical solutions, at an affordable price. To reach these quality standards, an important number of simulations using CODYRUN were made on each component of the building in order to quantify the thermal and energetic impact of each technical solution on the thermal comfort within the building. Our approach consisted of the study of typical dwellings, selected as the most representative of the type of accommodation built in the Reunion Island, in terms of architecture and building materials. We have selected two typical individual dwellings (one is a light structure the other a heavy structure) and a dwelling in a block of flats, as displayed in fig. 5.

*Fig. 5: Typical dwelling*

The simulations were carried out on the constituent components (roof, walls, windows) and on natural ventilation, in a way to estimate the influence of each ECODOM prescriptions, in terms of thermal comfort and energy loads performances. These simulations, their analysis and the synthesis of the results have been presented in a research report (in French), available from the authors or in [30]. With all the technical solutions of each component of the outer structure conformed to, in a second phase, we compared the existing dwelling which is on overall badly designed (bad solar protection, insufficient ventilation, etc...), to a well designed dwelling adhering to the technical solutions that had been found during the first phase.

*Fig. 6 : Typical dwelling results*

The simulations carried out led to the definition of high performance passive technical solutions for each comprising part of the structure and likewise a minimum ratio to optimise the natural ventilation. At the present date, (during 1997), three experimental operations, involving collective housing, at Reunion Island, have been given the seal of standard quality, standards which were a result of the simulations.
One is shown in figure 7 below.

*Fig. 7 : ECODOM Operation, "La Découverte", 40 dwellings, 1997.*

An experimental follow up of the dwellings will be organised for the first ECODOM operations in order to validate experimentally the impact of the passive cooling solutions on the comfort of the occupants. This follow up is important, as the setting up of the ECODOM standard will be the first step towards the setting up of thermal regulations in the French overseas departments, by the year 2000.

V) **THE CASE OF CODYRUN CONSIDERED AS A RESEARCH TOOL**

Generally, HVAC systems and especially domestic air conditioners, are considered only in either a very simplified way, or in an over elaborated one. The originality of our work is the multi-model approach in the modelling of air conditioning systems (in this case, we have studied the modelling of a split-system) and the integration into the calculation code CODYRUN. The aim is to compare an hourly time lapse model to a reduced time lapse model enabling the consideration of the regulation and the start-up of the system. An experimental validation was achieved in order to compare the accuracy of each defined model [31].The data acquisition system and the test cell are located at the University of Reunion Island.

Three modelling levels of air-to-air residential heat-pumps have been defined and integrated in the thermal building simulation software. The first model (model n°0) is to supply the demand of hourly cooling loads and electric demand with an ideal control loop (no delay between the solicitations and the time response of the system) . The second model (model n°1) takes into account the on/off cycle and the control processing. The third model (model n°2) is based on a linear model determined from the manufacturers data. Dynamic effects are taken into consideration in the last two models through a single time constant model [32]. The time step is reduced to one minute. The outputs are the total, sensible and latent cooling capacities at each time step. These models are linear, as the existing code and therefore do not increase the calculation time.

The first comparisons between the measurements and simulations show that a dynamic simulation with shorter time steps than one hour give much better results for the estimation of the electric consumption (see figure 8)

*Fig.8 : Electric power - comparison of measurements and simulations*

The system used for the experimental measurements is in fact oversized regarding the size of the cell. Therefore the system never reaches a permanent regime and has a low fractional on-time (16 %). Thus, the reduced time lap models give the best results as the start-up regime is the over-riding factor in relation to the permanent state. Model n°2 gives the best results regarding the energy consumption and the average COP over a day. In contrast, the hourly model under estimates the daily energy consumption by 60 % (see table 2).

*Table 2 : One hour time lap results*

The interest of the multimodel account of air handling systems is that the short time lap models could be used in a first simulation to estimate the COP as a function of the fractional ontime. In a second phase, a correction factor could be applied to the hourly model in order to determine the cooling capacity needs and the electricity consumption of a building during a long period. This will allow a better accuracy of the hourly model without sacrifying the calculation time.

**CONCLUSION**

For the building system, the multi model approach enables first and foremost the availability of various models in order to carry out the simulations. Thus, during a buildings history (conception, realisation and exploitation), a unique model does not meet the needs (thermal) of the user. However the selection of the models envisaged must be carried out in accordance to the time constraints linked to the numerical procedures and the availability of powerful computers. Overall the duration of the simulation periods varies from one day, for a particular phenomena, to a whole year, for the analysis of energy consumption. At the moment, the need for results at a certain deadline and the accuracy demanded, leads the designers to make great compromises. Thus, this demand for precision simulation and calculation time restraints leads to a selective use of the models.

The multiple model approach favours a movement towards a tool adapted to different objectives of the building thermal simulation. Further work must follow this logical step towards multi-user as regards the design and research aspect.

**Acknowledgements**

The financial contribution of E.D.F. (Electricité De France) and Reunion Island delegation of A.D.E.M.E. (Agence De l'Environnement et de la Maîtrise de l'Energie) to this study is gratefully acknowledged.

# FIGURES

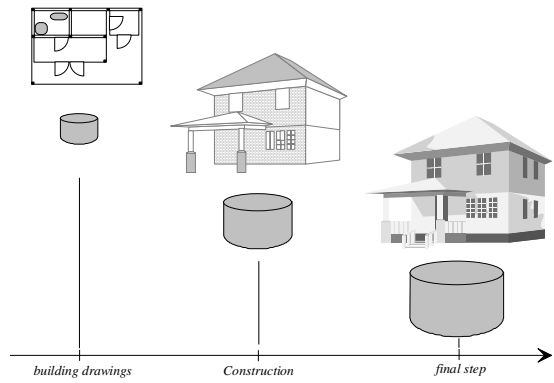

*Fig. 1 : Building database evolution*

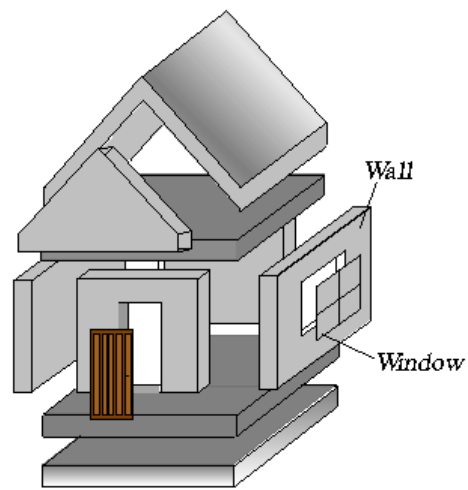

*Fig. 2 : Component assembly of a building*

*Fig. 3 : Building description window*

*Fig. 4 : External convection model window*

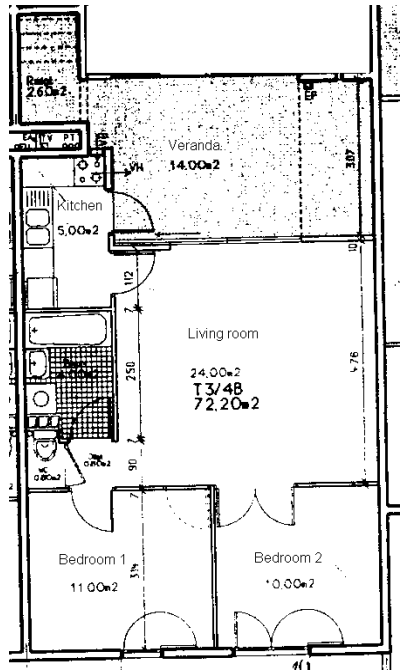

*Fig. 5: Typical dwelling*

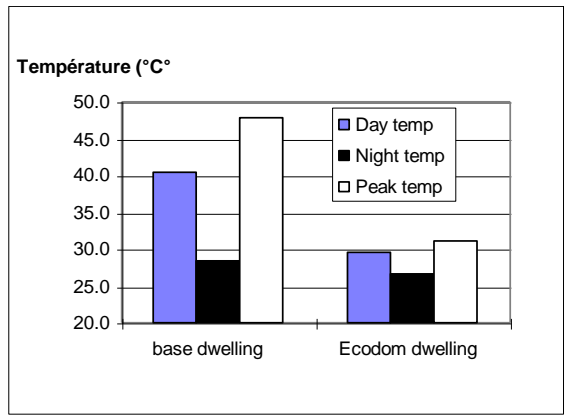

*Fig. 6 : Typical dwelling results*

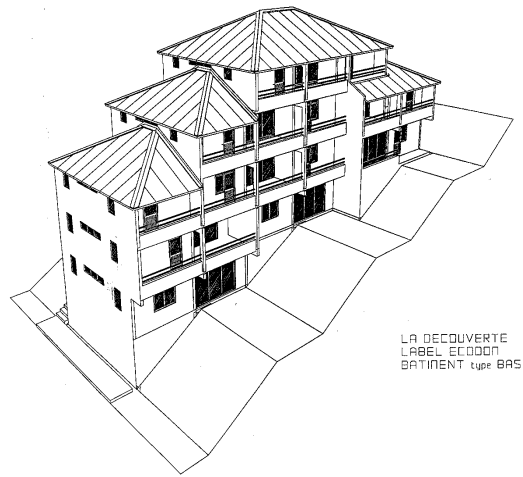

*Fig. 7 : ECODOM Operation, "La Découverte", 40 dwellings, 1997.*

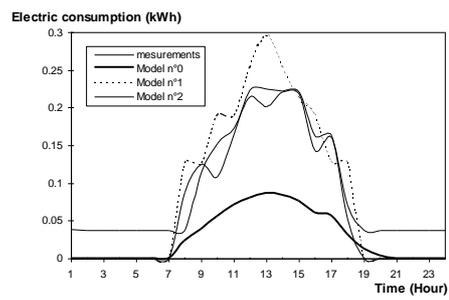

*Fig.8: Electric power - Comparison of measurements and simulations*

# TABLES

| Entity \ Model | Building | Zone | Components |
|---|---|---|---|
| Number of zones | x | | |
| Airflow transfer | x | | |
| Sky temperature | x | | |
| Outdoor convection | x | | |
| Diffuse reconstitution | x | | |
| Indoor convection | | x | |
| Indoor long wave radiation | | x | |
| Indoor short wave repartition | | x | |
| HVAC system | | | x |
| Heat conduction | | | x |

*Table 1 : model/entity allocation*

| | Daily Energy Consumption (kWh) | Relative difference | Mean COP |
|---|---|---|---|
| **Measurements** | 1.68 | | 1.6 |
| **Model n°0** | 0.68 | - 60% | 2.6 |
| **Model n°1** | 2.09 | + 25% | 1.22 |
| **Model n°2** | 1.76 | + 5% | 1.48 |

*Table 2 : One hour time lap results*